\documentclass[showpacs,prl,floatfix,twocolumn,amsmath,a4]{revtex4}

\usepackage{graphicx}

\begin{document} 

\title{High-$T_c$ ferrolectricity emerging from magnetic degeneracy in cupric oxide}

\author{Gianluca Giovannetti$^{1,2}$, Sanjeev Kumar$^{3}$, Alessandro Stroppa$^{4}$, Jeroen van den Brink$^{3}$, Silvia Picozzi$^{1}$, Jos\'e Lorenzana$^{2}$}

\address{
$^1$ Consiglio Nazionale delle Ricerche CNR-SPIN L'Aquila, Italy \\
$^2$ISC-CNR, Dipartimento di Fisica, Universit\`a ``La Sapienza'',
Piazzale Aldo Moro 5, 00185 Roma, Italy \\
$^3$Institute for Theoretical Solid State Physics, IFW Dresden, 01171 Dresden, Germany \\
$^4$CNISM- Department of Physics, University of L'Aquila,
Via Vetoio 10, 67010 Coppito, L'Aquila, Italy
}

\begin{abstract}
Cupric oxide is multiferroic at unusually high temperatures. From
density functional calculations we find that the low-$T$ magnetic phase
is paraelectric and the higher-$T$ one ferroelectric, with a size and
direction of polarization in good agreement with experiment. By
mapping the {\it ab initio} results onto an effective spin model we
show that the system has a manifold of almost degenerate ground states. 
In the high-$T$ magnetic state non-collinearity and inversion
symmetry breaking stabilize each other via the Dzyaloshinskii-Moriya
interaction. This leads to a novel mechanism for multiferroicity, with
the particular property that non-magnetic impurities enhance the
effect. 
\end{abstract}

\date{\today} 

\pacs{71.45.Gm, 71.10.Ca, 71.10.-w, 73.21.-b} 

\maketitle
 
In multiferroics the simultaneous presence of electric and magnetic ordering is particularly intriguing when the magnetic ordering triggers the ferroelectric polarization, as was observed for the first time by Kimura and coworkers in TbMnO$_{3}$~\cite{Kimura0}. Since then, several so-called type-II multiferroics~\cite{Brink2008} have been discovered in which \textit{magnetic order causes ferroelectric order}.  Although plenty of potential applications are envisioned, in random access memory devices for instance, the small values of the induced polarization as well as a low transition temperature in most type-II multiferroics hinder practical applications.  The very recent discovery that cupric oxide (CuO) is a type-II multiferroic with a high antiferromagnetic transition temperature $T_{N}$ of 230 K changed this situation drastically and opened the perspective to room-temperature multiferroicity~\cite{Kimura, Mostovoy}. The discovery is even more intriguing considering that CuO is closely related to the family of copper-oxide based materials displaying  High-$T_c$ superconductivity.  

From a theoretical point of view, the microscopic mechanism of multiferroicity in CuO is not clear yet, particularly
because its type-II behavior is apparently not a groundstate property:
it is only present at finite temperatures, between $\sim$210 and 230
K, disappearing above and below. Here we clarify the mechanism for the
observed finite temperature multiferroicity in CuO.   

To elucidate this point we have investigated the electronic structure
of CuO with density functional calculations for the different
magnetically ordered phases. These calculations confirm, as we will
see, the presence of magnetically induced ferroelectricity in CuO and
we find a polarization that agrees with experiment. A subsequent
investigation of the stability of magnetic phases at finite
temperatures using classical Monte-Carlo simulations shows that the
experimental ground state at low temperature can be well understood by
mapping the magnetic interactions onto a Heisenberg Hamiltonian with
{\it ab initio} derived exchange constants. The microscopic model
shows that the ground state is almost degenerate so phase selection
occurs through small terms. We show that 
multiferroicity in CuO arises from a new mechanism in which
spin canting and polarization mutually stabilize each other, crucially
involving phase selection through the Dzyaloshinskii-Moriya interaction. 
\begin{figure}
\centerline{\includegraphics[width=0.70\columnwidth,angle=0]{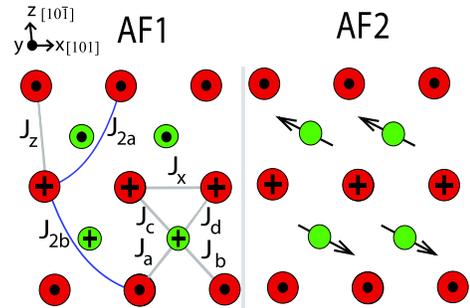}}
\caption{(Color online) Schematic view of AF1 and AF2 magnetic states.
  Cu ions belong to different plane along $y=0$ (red/big
  circle, termed even plane in the text) and
  $y=1/2$ (green/small circle, termed odd plane in the text). 
Dot (cross) refers to spin pointing along the positive (negative) $y$
axis.  }
\label{fig1}
\end{figure}

We have studied the electronic structure of CuO by performing
calculations using the PAW method as implemented in
VASP~\cite{VASP}. To take into account the Coulomb interactions
between the Cu $3d$ electrons we employ
SGGA+U~\cite{PW91,PAW+U,Dudarev} and hybrid functional
(HSE)~\cite{hse} schemes for $U_{eff}=U-J_H$ ranging between 3.5 and
7.5 eV, and fraction of Hartree-Fock (HF) exchange ($\alpha$) between
$0.15$ and $0.25$. In the experimental structure of CuO (C2/c space
group No. 15), Cu ions are arranged as corner and edge-sharing
square-planar CuO$_4$s, in which Cu ions lie in a plane formed by
oxygen neighbors forming O-centered tetrahedra~\cite{Asbrink}. The low
temperature magnetic structure (AF1) with Cu magnetic moments aligned
collinearly along the $y$ axis and ordered antiferromagnetically along
$z$ and ferromagnetically along the $x$ 
direction (see Fig. \ref{fig1}) is found to be the ground state in
agreement with the neutron diffraction study\cite{Forsyth} and other
ab-inito calculations~\cite{Rocquefelte,Filippetti}. 
The high N\'eel temperature suggests the presence of strong exchange
interactions~\cite{Yang,Ain,Boothroyd} and  the quasi-one-dimensional
antiferromagnetism of CuO is suggested by
experiments\cite{Shimizu,Jung}. The energy gap of 1.4
eV~\cite{Ghijsen} is recovered using $U_{eff}$=5.5 eV or an
exact-exchange fraction as $\alpha=0.15$~\cite{Rocquefelte}. The Cu
and O spin moments are 0.6 $\mu_{B}$ and 0.1 $\mu_{B}$ respectively,
in good agreement with the results of neutron diffraction
experiments~\cite{Forsyth}.  

\begin{table}
\begin{center}
\begin{tabular}{|c|c|c|c|c|c|c|c|}
\hline
 & $J_z$ & $J_x$ & $J_{2a}$ & $J_{2b}$ & $J_a = J_d$ & $J_b = J_c$ & $J_y$ \\
\hline
$U_{eff}$=5.5 &107.76 & -15.76 & 6.89 & 16.18 & 7.98 & 15.82 & -21.48  \\
$\alpha$=0.15 &120.42 & -24.33 & 4.99 & 14.27 & 4.19 & 13.17 & -23.02 \\
\hline
\end{tabular}
\label{table2}
\caption{Exchange coupling parameters (meV) calculated within SGGA+U and
  hybrid functional calculations. The structure allows for $J_a\ne
  J_d$ and  $J_b\ne J_c$ but we take them equal for simplicity. This
  is inessential for our conclusions. We keep the same notation of reference
~\cite{Rocquefelte}.}
\label{table1}
\end{center}
\end{table}

We analyze the exchange interactions using the Heisenberg Hamiltonian 
$
H_M= \sum_{ij}  J_{ij} ~ {\bf S}_i \cdot {\bf S}_j   
$.
The parameters $J_{ij}$, reported in Table \ref{table1}, are
calculated from total energy differences of different magnetic
configurations within a unit cell containing 32 Cu sites. The
strongest interaction is $J_z\sim 100$ meV  in good agreement with the
optical experiments of Ref.~\cite{Jung} and the neutron data of
Ref.~\cite{Boothroyd}.  
The magnetic structure within a constant-$y$ plane can be viewed as
chains running along $z$ with dominant interaction $J_z$ and moderate interchain interactions $J_x$, $J_{2a}$ and $J_{2b}$. Hereafter we will term the planes with $2y$ odd (even) as odd (even) planes.  $J_y$ couples planes of the same kind (separated by  $\Delta y=\pm 1$). Finally $J_a$, $J_b$, $J_c$ and $J_d$ couple nearest neighbor (nn) planes of different kind. 

Remarkably the classical coupling energy among unlike planes vanishes
when the chains are assumed to be aligned AF. Indeed $J_y$ favors a ferromagnetic staking of planes of the same kind. A close examination of the structure shows that the pattern of couplings among the plane at 
$y=1/2$ and the plane at $y=1$ is identical to the one of
Fig.~\ref{fig1} but with the exchanges $J_a\leftrightarrow J_d $ and
$J_c\leftrightarrow J_b$. It is easy to check that this symmetry makes
the classical energy of the model independent of the angle between the
magnetization in even and odd planes, and the ground state is
infinitely degenerate. This degeneracy will play an important role for
the multiferroic mechanism. 

Relaxation of charge and lattice and small
anisotropies favor a collinear structure. Indeed within DFT
calculations, including spin-orbit (SO) coupling, we find that AF1 is the ground
state with an energy gain of 2.2 meV per Cu atom respect to the
non-collinear AF2 structure shown in Fig.~\ref{fig1}. The easy axes is
found to be along the $y$ direction.  
   

In CuO, an incommensurate phase with magnetic modulation vector
Q=$(0.506,0,0.517)$ has been reported at the temperatures where the
ferroelectricity is observed~\cite{Forsyth,Yang,Kimura}. In this phase
spins order along the $y$ axis
and in the $xz$ plane with spins on even and odd planes
perpendicular to each other. This is shown schematically in the right
panel of Fig.~\ref{fig1} where the small incommensuration of the
structure has been neglected.    
Experimentally, a small electric polarization $P$
($\sim$0.01 $\mu C/cm^2$) is found along $y$ axis\cite{Kimura}.

On a first sight one would expect that the incommensurate spiral is
crucial to obtain a finite polarization, as in the standard cycloid
scenario\cite{Cheong,Kenzelmann,Mostovoy}. However the 
commensurate state closest to the incommensurate spiral, labeled AF2 in  
Fig.~\ref{fig1}, has spin canting which
can produce a finite polarization. Indeed,  taking into
account spin-orbit coupling, we evaluate the electronic contribution to
the polarization $P$ using the Berry phase (BP) 
method~\cite{BerryPhase} on the {\em commensurate} AF2 state 
and we obtain $P_{AF2} \sim$ 0.02 $\mu
C/cm^2$ along $y$  axis, in overall good agreement with the
experimental value.  Thus the incommensurate state is not crucial but
canting clearly is.  The perpendicular configuration ensures that AF2
state has maximal spin current  ${\bf j}_{1,2}\equiv \langle {\bf
  S}_1\times {\bf S}_2\rangle$ among nn planes of different kind. We
will show that this is a fingerprint of the proposed scenario. 

\begin{figure}
\centerline{\includegraphics[width=0.94\columnwidth,angle=0,clip=true]{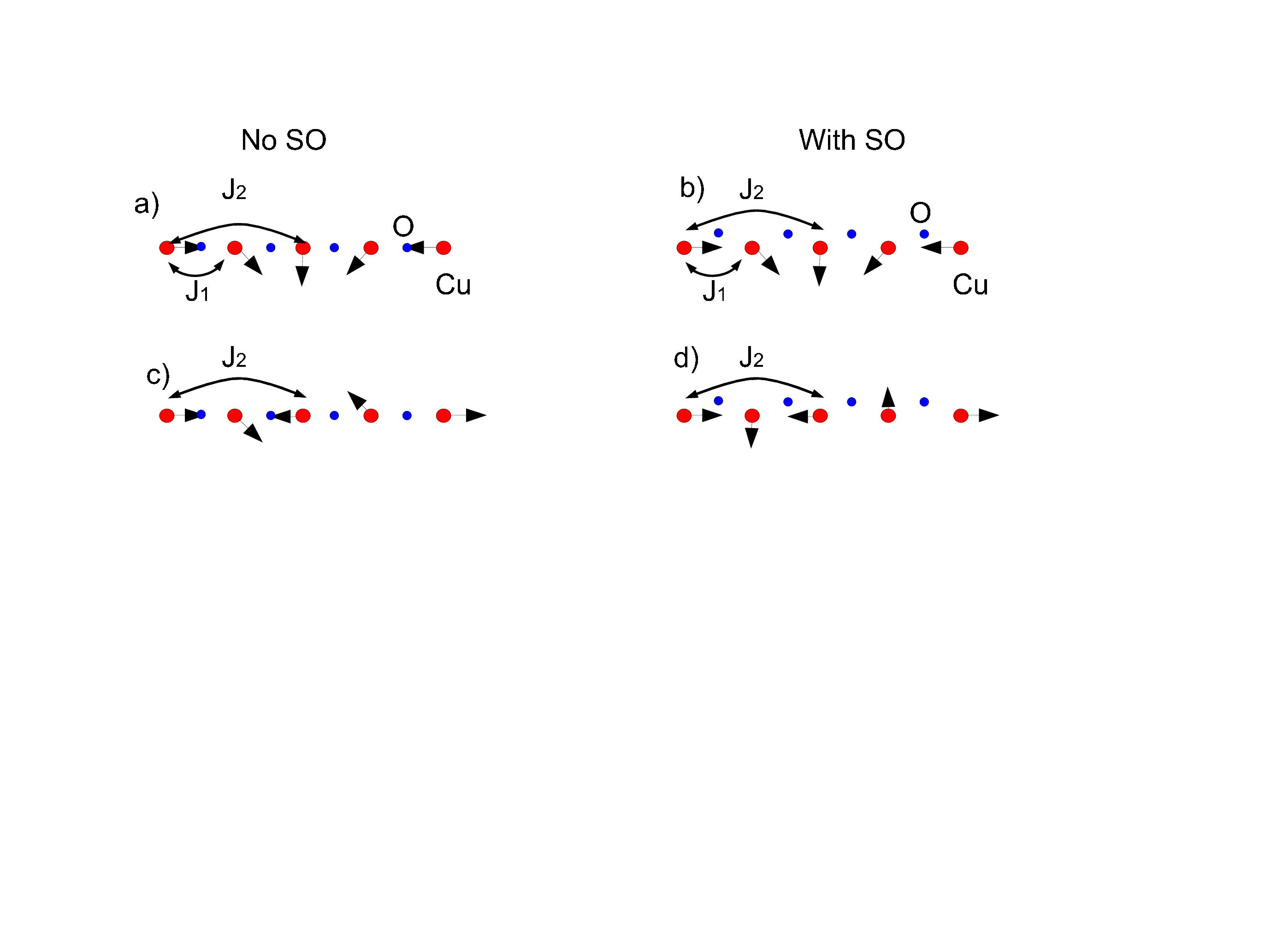}}
\caption{(Color online) (a) Schematic mechanism of multiferroic effect
  in 1D.  a) and b) standard cycloid scenario as in Ref.~\cite{Cheong}.  In c) $J_1=0$ and the
  system separates into two interpenetrating sublattices. The angle
  between the magnetic moments of the different sublattices is
  undefined. d) SO interaction stabilizes both the angle and the finite
  polarization. 
 }
\label{schematic}
\end{figure}

While canting and spin orbit coupling are standard ingredients of the
cycloid scenario causing the
multiferroicity \cite{Cheong,Kenzelmann,Mostovoy}, the situation in CuO is 
subtly different. In order to explain the difference and similarities we illustrate the two mechanisms
in the one-dimensional (1D) model \cite{Cheong} 
depicted in Fig.~\ref{schematic}. Consider an
hypothetical  Cu-O chain with Hamiltonian $H=H_M+H_{DM}+H_{E}$. 
For the purely magnetic part $H_M$ we  assume there is a nearest neighbor AF
interaction $J_1$ and a next nearest neighbor AF interaction
$J_2$. The Dzyaloshinskii-Moriya (DM) interaction~\cite{Dzyaloshinskii,Moriya}
 and elastic contributions read:
\begin{eqnarray}
  \label{eq:h1d}
H_{DM}&=&\sum_n \lambda ({\bf u}_{n+1/2} \times {\bf e}_{n,n+1}) . ({\bf S}_n
\times {\bf S}_{n+1}),\\
H_{E}&=&\sum_n\frac{k}2|{\bf u}_{n+1/2} |^2.
\nonumber
\end{eqnarray}
Here ${\bf u}$ are the oxygen displacements and ${\bf e}_{n,n+1}$ is a
unit vector joining nearest neighbors atoms. 
Treating the spin classically for 
$J_2>J_1/4$ one finds that the ground state is a spiral with a pitch
angle given by $\cos\theta=-J_1/4J_2$ and a finite spin current 
${\bf j}=\langle {\bf S}_n \times {\bf S}_{n+1} \rangle$.  SO
interaction is not necessary to stabilize this state [Fig.~\ref{schematic}(a)].
In the presence of SO coupling the free energy
per site due to
uniform displacements $u$ of oxygens, perpendicular to the chain and to ${\bf
  j}$ is given by $\delta F_{DM}=\lambda u j + \frac{k}2
u^2$. Minimizing, one obtains a polarization $P=\delta q  u 
 = -\delta q j/ k$ [Fig.~\ref{schematic}(b)] due to the difference in
 charge $\delta q$ of the two ions.

In the case $J_1=0$ [(c) and (d)] the system separates into two interpenetrating
sublattices analogous to the odd and even planes of CuO. 
Without SO coupling the  angle between the magnetic
moments of the different sublattices is arbitrary and the ground state
is infinitely degenerate (c). 
This degeneracy can be broken by the DM interaction. 
The free energy per site can be expanded  as: 
$\delta F= \frac12 \chi_{jj}^{-1} j^2 + \lambda u j + \frac{k}2
u^2$. Here $\chi_{jj}$ is a spin current  susceptibility defined for
$H_M$ alone. 
$\delta F$ has to be minimized with respect to {\em both} $j$ and $u$ since 
$H_M$ does not determine the spin current.  
Minimizing with respect to $u$ one obtains 
$\delta F=\frac12 (\chi_{jj}^{-1}-\lambda^2/k) j^2$. 
When
\begin{equation}
\label{eq:condition}
 \chi_{jj} \frac {\lambda^2}k> 1
\end{equation}
it is convenient to maximize the
spin current $j\propto \sin \phi$  where $\phi$ is the angle between
the magnetization in different sublattices.  Thus the energy  acquires
a term  $\delta F \propto - \sin^2 \phi $ which favors perpendicular
magnetic moments on nearest neighbor sites [Fig.~\ref{schematic}(d)].
At the same time this gives rise to a finite polarization as for the
cycloid mechanism.  Notice however that the spontaneous breaking of
symmetry and the spin canting drive each other unlike the cycloid
scenario where the spin canting is driven by the magnetic
Hamiltonian. 
From Eq.~(\ref{eq:condition}) we see that the multiferroic effect
is favored by strong SO coupling,  soft lattices and a large
spin current susceptibility. 
This effect competes with thermal and quantum fluctuation which
favor a collinear configuration, according to the order by disorder 
mechanism, and tend to suppress  $\chi_{jj}$\cite{Henley}. 
Also coupling to other lattice distortions
will favor a collinear state 
trough effective biquadratic terms in the Hamiltonian [$\sim ({\bf S}_i.{\bf S}_j)^2$] and suppress $\chi_{jj}$.  

The 1D mechanism can be easily generalized to CuO by replacing
each magnetic site in Fig.~\ref{schematic} by a constant-y plane of Cu atoms
shown in Fig.~\ref{fig1}.

In order to estimate $\chi_{jj}$  for the CuO structures we perform classical Monte-Carlo
simulations on the following 3D spin Hamiltonian:
$ H = H_M  + \sum_{(ij)} {\bf D}\cdot {\bf j}_{i,j}$.
Here, the first term is the magnetic Hamiltonian of CuO
and the second term describes the linear coupling
of the spin current among unlike planes to an auxiliary external field ${\bf D}$. 
Summation index $(ij)$ represents inter-plane nearest neighbor bonds
indicated as $J_a$, $J_b$, $J_c$ and $J_d$ in Fig. \ref{fig1}. 
We take ${\bf D}$  directed along $x$-axis for all pairs
in the unit cell. Rather than measuring the spin current we compute
the sum of the relevant components which is proportional to the polarization,
${\bf p} = \sum {\bf e}_{ij} \times \langle {\bf S}_i \times {\bf S}_j
\rangle$. The desired susceptibility is given by $\chi_{jj}=p_y/D_x$
in the limit of vanishing $D_x$.
When $\chi_{jj}$ is large and the condition Eq.~(\ref{eq:condition}) is
satisfied a spontaneous $D$ and polarization will stabilize each other
as explained above. 
Notice that $D$ breaks inversion symmetry in the 
system whereas the high temperature structure of CuO has inversion so
there is no ``permanent'' $D$ in the high temperature phase. Of course
there will be DM couplings at high-temperatures in the CuO structure
 but those preserve inversion symmetry and are not related to the
 appearance of the polarization. For simplicity we neglect these
 latter couplings.  

To lift the degeneracy between AF1 and AF2 in favor of the former we introduce 
a weak  anisotropy in 
the Heisenberg exchange term. This is done by replacing
$J_z {\bf S}_i \cdot {\bf S}_j$ by  $J_z (S^x_i S^x_j +
(1+\gamma)S^y_i S^y_j + S^z_i S^z_j)$ in $H_M$.  Other terms like the
biquadratic contribution are expected to have a similar effect.  
We use $\gamma= 0.02$ which translates into an anisotropy energy of
$\sim \gamma J_z = 2.15$ meV. 

\begin{figure}
\centerline{\includegraphics[width=0.90\columnwidth,angle=0,clip=true]{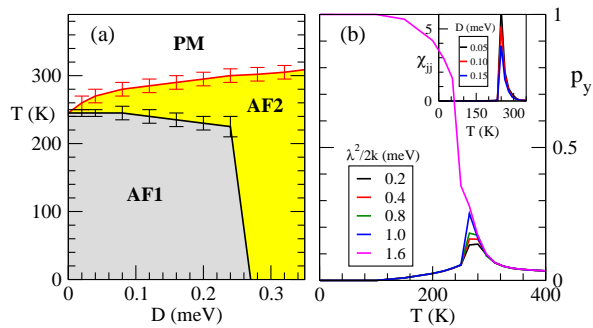}}
\caption{(Color online) (a) Phase diagram of the magnetic model of CuO
  with an external field coupling linearly with the spin current. 
(b) Polarization of CuO model with the addition of a biquadratic
spin-current term obtained eliminating the lattice degrees of freedom
in the DM coupling. Curves are labeled by the value of $\lambda^2/(2k)$. 
The inset shows the susceptibility computed as
$p_y/D$. Within linear response one should  take the
limit $D \rightarrow 0$ corresponding to the upper curves.}
\label{fig3}
\end{figure}

We employ a classical Monte Carlo (MC) technique to explore the
competition between different magnetic states at zero and finite
temperatures. Given that the CuO is a system with spin 1/2, the
quantum effects in this system are unavoidable.  
Nevertheless the interesting transitions occur at high temperatures where it is
safe to assume a classical renormalized regime\cite{Chakravarty}. 
In order to simplify the Monte Carlo computation we consider only 4
possible states  at 90$^0$ for the spin variables.   
 Figure \ref{fig3}(a) shows the phased diagram in the $T$-$D$ plane.
In the absence of the external field $D$, the system undergoes a
transition from a paramagnetic (PM) to an AF1  
state with $T_N \sim 250K$. Presence of a small $D$ opens a narrow
window near $T_N$ where AF2 is stabilized.  A large external field
eventually drives the groundststate to be AF2 for $D = \gamma J_z/8
\sim 0.27$ meV . 
The inset of Fig.~\ref{fig3}(b) shows the susceptibility computed as
the ratio  $p_y/D_x$ for small $D_x$. Remarkably a strong peak
appears around the PM to AF1 transition which will favor a
spontaneous polarization of the system. 

In order to check the mechanism we again consider the 3D CuO model with the terms
$H_{DM}$ and $H_{E}$ analogous to the 1D model. Assuming classical
lattice displacements they can be integrated out of the partition
function leading to a quadratic
effective interaction among spin currents so $H_{DM}+H_{E}$ is
replaced by
$H_{DME}= -(\lambda^2/2k)\sum_{(ij)} ({\bf S}_{i}\times{\bf S}_{j})^2$.
Fig.~\ref{fig3}(c) shows the spontaneous polarization $p$ as a function
of temperature for the model defined by $H=H_M+H_{DME}$. As expected
one finds that, close to the PM to AF1 transition, $H_{DME}$ induces a
phase with 
broken inversion symmetry and a spontaneous polarization as seen in
the experiment. The peak in the polarization is  very similar to the 
experimental observations of a finite electrical polarization between
230K and 213K~\cite{Kimura}. If the parameter $\lambda^2/(2k)$ is made
too large ($>1$ meV) the ferroelectric phase extends to zero
temperature. 

According to our Monte Carlo simulations the main reason for the
spontaneous polarization is a strongly enhanced spin-current susceptibility
close to the AF1-PM transition. While there is of course a divergent
staggered susceptibility when approaching the AF1-PM phase transition
the enhancement of the unrelated spin-current susceptibility is not
trivial. This is similar to the physics of quantum
critical points relevant to heavy fermion
compounds\cite{Monthoux} where close to the quantum transition
between a disordered an a magnetically order state a different order
appears (superconductivity) which can be attributed to an enhanced
pairing susceptibility. 

As mentioned above thermal and quantum fluctuations tend to suppress
$\chi_{jj}$ and the polarization. On the other hand disorder on the
magnitude of the magnetic moments will enhance the tendency to have
perpendicular orientations among the
sublattices and enhance the polarization\cite{Henley} which opens a
new way to engineer high-$T_c$ multiferroic
materials. Non-magnetic impurities, vacancies or
magnetic impurities with a 
different spin will lead to this effect.  In addition our results
suggest to search for other materials where two subsystems have
negligible interactions by symmetry but strong interactions within one
subsystem as a recipe to discover new manipulable multiferroics.

To summarize, our density functional calculations confirm the
magnetically induced ferroelectricity in CuO with polarization in
agreement with experiments.  By combining Monte-Carlo analysis with
the exchange constants derived by {\it ab initio} simulations we also
confirm the high $T_{N}$ of this compound. We explain the multiferroic
effect as arising from a new mechanism in which spin canting and
polarization mutually stabilize each other with a crucial role of
Dzyaloshinskii-Moriya interaction.  Our results open new routes for the
material design of multiferroics.

This work is supported by the European Research Council through the
BISMUTH project (Grant N. 203523) and IIT-Seed project NEWDFESCM. We
thank A. Boothroyd for a critical reading of the manuscript.  

During the completion of this manuscript  we became aware of
Ref.~\cite{Jin2010}, presenting results from {\it ab initio}
calculations similar to ours. The mechanism for multiferroicity that
we present here, however, is very different from the findings in
Ref.~\cite{Jin2010}.

\end{document}